\documentclass{article}

\PassOptionsToPackage{numbers,sort&compress}{natbib}

\usepackage[preprint]{neurips_2022}

\usepackage[utf8]{inputenc}
\usepackage[T1]{fontenc}
\usepackage{hyperref}
\usepackage{url}
\usepackage{booktabs}
\usepackage{amsfonts}
\usepackage{amsmath,amssymb}
\usepackage{nicefrac}
\usepackage{microtype}
\usepackage{xcolor}
\usepackage{graphicx}
\usepackage{multirow}
\usepackage[ruled,vlined,linesnumbered]{algorithm2e} 

\title{Test-Time Adaptation for Non-stationary Time Series:\\
From Synthetic Regime Shifts to Financial Markets}

\author{
  Yurui Wu,\ Qingying Deng,\ Wonou Chung,\ Mairui Li\\
  Department of Statistics \& Operations Research, UNC Chapel Hill\\
  Chapel Hill, NC, 27599\\
  \texttt{\{yuruiwu,doradeng,wchung,mairuili\}@email.unc.edu}
}

\begin{document}

\maketitle

\begin{abstract}
Time series encountered in practice are rarely stationary. When the data distribution changes, a forecasting model trained on past observations can lose accuracy. We study a small-footprint test-time adaptation (TTA) framework for causal time-series forecasting and direction classification. The backbone is frozen, and only normalization affine parameters are updated using recent unlabeled windows. For classification we minimize entropy and enforce temporal consistency; for regression we minimize prediction variance across weak time-preserving augmentations and optionally distill from an EMA teacher. A quadratic drift penalty and an uncertainty-triggered fallback keep updates stable. We evaluate this framework in two stages: synthetic regime shifts on ETT benchmarks, and daily equity and FX series (SPY, QQQ, EUR/USD) across pandemic, high-inflation, and recovery regimes. On synthetic gradual drift, normalization-based TTA improves forecasting error, while in financial markets a simple batch-normalization statistics update is a robust default and more aggressive norm-only adaptation can even hurt. Our results provide practical guidance for deploying TTA on non-stationary time series. The source code of this study is available at this \href{https://drive.google.com/drive/folders/11L2n1PZDQBolXU5Re1RxcfnWOD96t3FT?usp=drive_link}{Google Drive}.
\end{abstract}

\section{Introduction}
Real-world time series, such as financial prices or electricity loads, evolve over time. Structural breaks, policy changes, and macroeconomic shocks can shift both the level and the volatility of the process. During the COVID-19 crisis and the subsequent high-inflation period, market dynamics changed rapidly, and models trained on pre-crisis data became miscalibrated. Once the deployment distribution departs from the training distribution, minimizing training loss is no longer sufficient to guarantee test performance.

Test-time adaptation (TTA) addresses this issue by updating part of the model using unlabeled test inputs. For time series, a practical TTA scheme must be causal (no future data), parameter efficient, and robust to noisy updates. In this work we freeze the backbone and adapt only a small parameter set at deployment. When day $t$ arrives, we collect the most recent windows, apply weak time-preserving augmentations, and take a few gradient steps on unsupervised objectives. For classification we combine entropy minimization with temporal consistency; for regression we minimize augmentation variance and optionally distill from an exponential-moving-average teacher. We control the size of daily parameter changes and fall back to a lightweight batch-normalization statistics update when an uncertainty proxy is high.

Our goal is not to propose a new heavy architecture, but to understand when small-footprint TTA helps or hurts on non-stationary sequences. We make three contributions. First, we unify several TTA choices---no adaptation, batch-normalization statistics refresh, and norm-only adaptation---under one causal framework. Second, we evaluate this framework in two stages: synthetic shifts on ETT benchmarks, and real markets (SPY, QQQ, EUR/USD) split into pandemic, high-inflation, and recovery regimes. Third, we connect the results with econometric tools (Diebold--Mariano tests and Newey--West adjustments) and summarize practical recommendations for deploying TTA in streaming forecasting systems.

\section{Literature Review}
\label{sec:related}

\subsection{Test-time adaptation under distribution shift}
TTA minimizes an unsupervised objective at inference using the current test inputs, while most parameters are frozen. A central result is that tuning only normalization affine parameters $(\gamma,\beta)$ by entropy minimization can yield large gains under covariate shift \citep{wang2021tent}. This works because normalization re-centers and re-scales hidden activations to match the present input distribution, and the affine degrees of freedom are enough to correct many first/second-moment changes. In streaming scenarios, adaptation can accumulate error if the procedure is aggressive or if pseudo-targets are unstable. Prior work proposes stabilizers such as exponential-moving-average teachers and conservative update schedules \citep{wang2022cotta,yuan2023rotta}. We follow this line by using an EMA teacher for self-distillation in regression and by adding a quadratic drift penalty that discourages large inter-day parameter motion. 

Refreshing batch-normalization statistics at test time (recomputing means and variances without gradients) already improves robustness to small-batch and covariate shifts \citep{yang2022ttbn,schneider2020covariate}. This motivates our fallback: when an uncertainty proxy is high, we refresh statistics and skip gradient updates. Recent analyses add explanatory depth; for example, Kim and Lee et al.\ \citep{kim2024ttalinearity} show that TTA can increase linearity in intermediate layers, which yields more stable extrapolation on shifted inputs. For forecasting, test-time learning can be formulated with variance and consistency objectives that do not require labels \citep{christou2024ttl}. Adaptation in non-stationary environments can also be framed as representation alignment that maintains proximity between current and reference features \citep{zhang2024nonstationarytta}. Our consistency and variance objectives, together with EMA-teacher and drift control, are concrete realizations of these ideas for time series.

\subsection{Normalization and architectures for non-stationary sequences}
Non-stationarity often manifests as level/scale drift and time-varying seasonality, so normalization layers are a primary tool. Reversible instance normalization (RevIN) standardizes each sequence and then reverses the transform before output; it reduces train--test mismatch while preserving the forecasting target \citep{kim2022revin}. Adaptive normalization learns gates that respond to distributional changes \citep{liu2023san}. Frequency-aware normalization emphasizes multi-scale periodic structure and modulates features accordingly \citep{ye2024fan}. These methods act within the backbone. Our TTA instead acts on top of such defenses, allowing small affine corrections that follow the deployment stream, so the two approaches are complementary. On the architecture side, the Non-Stationary Transformer incorporates evolving statistics directly into attention and residual paths \citep{liu2022nonstationary}. For our hosts we adopt light yet competitive models—multi-scale residual  Temporal Convolutional Networks (TCN) \citep{bai2018tcn} and compact Transformers—as well as PatchTST and TimesNet as references \citep{nie2023patchtst,wu2023timesnet}. We select small widths (64--128) and few layers (2--4) to ensure that test-time updates touch only thousands of parameters, keeping latency negligible while preserving accuracy.

\subsection{Regime-wise evaluation and econometric significance}
Statistical performance should be interpreted by regime. Markov-switching models explain how coefficients and volatility change across states \citep{hamilton1989regimes}, and survey evidence supports regime-dependent predictability in financial markets \citep{ang2011regime}. We therefore report rolling curves and regime-wise summaries. To compare forecast accuracy we use the Diebold--Mariano test with HAC variance estimation \citep{diebold1995dm,newey1987nw}. When we evaluate many variants, standard multiple-comparison corrections may be anti-conservative; Reality Check and SPA address data-snooping by bootstrapping the maximum performance statistic across models \citep{white2000reality,hansen2005spa}. Our significance reporting is aligned with these tools.

\section{Problem Setup}
\label{sec:setup}
Let $\{x_t\}_{t=1}^T$, $x_t\in\mathbb{R}^d$ be a multivariate series. We form an input window $X_t=[x_{t-L+1},\dots,x_{t}] \in \mathbb{R}^{L\times d}$ and study two tasks. 

\begin{enumerate}
\item Classification: predict the direction for day $t{+}1$; let $y_{t+1}\in\{0,1\}$ denote down or up, and let $\hat{p}(X_t)\in[0,1]^2$ be the predicted probability. We track accuracy, F1, AUC, direction accuracy, and expected calibration error (ECE), and draw reliability diagrams.
\item Regression: predict next-day log-return $r_{t+1}=\log p_{t+1}-\log p_t$ or $H$-day cumulative log-return $r_{t+1:t+H}=\sum_{h=1}^{H} r_{t+h}$; we report MAE, RMSE, and $R^2$. 
\end{enumerate}

The deployment stream $\{P_t\}$ is non-stationary relative to training $P_{\text{train}}$, with shifts including location-scale drift, noise inflation, and structural changes. Synthetic generators in Appendix~\ref{app:shifts} implement these. For daily equity and FX series we use a unified split: train 2000--2016, validation 2017--2019, test 2020--2025. All scalers are fit on training only, and we avoid exogenous variables to prevent leakage. ETT benchmarks follow their standard splits.

\section{Method}
\label{sec:method}
We train a backbone $f_\theta$ on the training split with supervised losses. At test time we adapt a small parameter set on unlabeled windows from the current stream; the backbone is frozen.

\subsection{Backbones and updateable units}
We consider a multi-scale residual TCN and a compact Transformer (2--4 layers, width 64--128). GRU/LSTM serve as references. At test time, we compare the following options:
\begin{itemize}
    \item \textbf{no\_tta}: no test-time adaptation, frozen model.
    \item \textbf{bn\_stats}: refresh batch-normalization (BN) means and variances using the current batch, no gradients.
    \item \textbf{norm\_only}: update only normalization affine parameters $(\gamma,\beta)$.
\end{itemize}

\subsection{Unsupervised objectives}
Let $\mathcal{B}$ be the batch of recent unlabeled windows. For classification we use entropy and consistency:
\begin{align}
\mathcal{L}_{\mathrm{ent}}(\theta) &= \frac{1}{|\mathcal{B}|}\sum_{X\in\mathcal{B}} \Big(-\sum_{c} \hat{p}_c(X)\log \hat{p}_c(X)\Big), \label{eq:ent}\\
\mathcal{L}_{\mathrm{cons}}(\theta) &= \frac{1}{|\mathcal{B}|}\sum_{X\in\mathcal{B}} \big\|\hat{p}(X)-\hat{p}(T(X))\big\|_2^2, \label{eq:cons}
\end{align}
where $T$ is a weak time-preserving transform (Section~\ref{sec:aug}). For regression we use variance minimization and EMA-teacher self-distillation:
\begin{align}
\mathcal{L}_{\mathrm{var}}(\theta) &= \frac{1}{|\mathcal{B}|}\sum_{X\in\mathcal{B}} \mathrm{Var}\big(\{\hat{y}(T_k(X))\}_{k=1}^K\big), \label{eq:var}\\
\mathcal{L}_{\mathrm{sd}}(\theta,\tilde{\theta}) &= \frac{1}{|\mathcal{B}|}\sum_{X\in\mathcal{B}} \big\|\hat{y}(X)-\tilde{y}(X)\big\|_2^2,\qquad \tilde{\theta}\leftarrow \rho\tilde{\theta}+(1-\rho)\theta. \label{eq:sd}
\end{align}

Drift control: we penalize inter-day change in the small parameter set,
\begin{equation}
\mathcal{L}_{\mathrm{drift}}(\theta^{(t)},\theta^{(t-1)})=\gamma\,\big\|\theta^{(t)}-\theta^{(t-1)}\big\|_2^2. \label{eq:drift}
\end{equation}
Total objective: for classification,
\begin{equation}
\mathcal{L}(\theta)=\alpha\,\mathcal{L}_{\mathrm{ent}}+\beta\,\mathcal{L}_{\mathrm{cons}}+\mathcal{L}_{\mathrm{drift}},
\label{eq:total_cls}
\end{equation}
and for regression,
\begin{equation}
\mathcal{L}(\theta)=\alpha\,\mathcal{L}_{\mathrm{var}}+\beta\,\mathcal{L}_{\mathrm{sd}}+\mathcal{L}_{\mathrm{drift}}.
\label{eq:total_reg}
\end{equation}
Appendix~\ref{app:deriv} provides derivations and gradients for norm-only parameters, including proximal interpretations of the drift penalty and the variance-reduction role of the EMA teacher.

\subsection{Uncertainty-triggered fallback}
We compute an uncertainty proxy on $\mathcal{B}$. For classification, we use mean entropy $u_t=\frac{1}{|\mathcal{B}|}\sum_X H(\hat{p}(X))$. For regression, we use mean augmentation variance $u_t=\frac{1}{|\mathcal{B}|}\sum_X \mathrm{Var}(\{\hat{y}(T_k(X))\}_k)$. We set $\tau$ as the 80th percentile of $u_t$ on validation; 70\% and 90\% are ablated. If $u_t>\tau$, we refresh BN statistics and skip gradient updates.

\subsection{Causality-preserving augmentations}
\label{sec:aug}
We use weak transforms that keep time order and avoid leakage: amplitude scaling ($1\pm 5\%$), Gaussian jitter with standard deviation $0.01$ times the training standard deviation, time jitter by $\pm 1$ step within the window, and time cutout masking up to 5 steps.

\subsection{Daily test-time adaptation algorithm}
Default hyperparameters are $W{=}64$, $S{=}5$, learning rate $10^{-4}$, and $\alpha=\beta=1$, $\gamma=10^{-3}$.

\begin{algorithm}[H]
\DontPrintSemicolon
\caption{Causal Test-Time Adaptation for Day $t$}
\label{alg:tta}
\KwIn{Frozen backbone $f_\theta$; updatable parameter set $\phi\subset\theta$ (norm-only BN affine parameters);\;
Window length $L$; context size $W$; steps $S$; learning rate $\eta$; uncertainty threshold $\tau$;\;
Loss type: classification (\eqref{eq:total_cls}) or regression (\eqref{eq:total_reg}); transforms $\{T_k\}$.}
\KwOut{Prediction $\hat{y}_{t+1}$ (or $\hat{p}_{t+1}$) for the next day.}
Build unlabeled batch $\mathcal{B} \leftarrow \{X_{t-W+1},\dots,X_t\}$.\;
Compute uncertainty $u_t$ (entropy or augmentation variance).\;
\If{$u_t > \tau$}{
  Refresh BN running statistics on $\mathcal{B}$; \Return $\hat{y}_{t+1} \leftarrow f_{\theta}(X_t)$.\;
}
\For{$s=1$ \KwTo $S$}{
  \uIf{classification}{
    Sample a weak transform $T$; compute $\mathcal{L}_{\mathrm{ent}}$ and $\mathcal{L}_{\mathrm{cons}}$.\;
    $\mathcal{L}\leftarrow \alpha\mathcal{L}_{\mathrm{ent}}+\beta\mathcal{L}_{\mathrm{cons}}+\gamma\|\phi-\phi^{\text{prev}}\|_2^2$.\;
  }
  \uElse{regression}{
    Sample $K$ transforms $\{T_k\}$; compute $\mathcal{L}_{\mathrm{var}}$ and optionally $\mathcal{L}_{\mathrm{sd}}$.\;
    $\mathcal{L}\leftarrow \alpha\mathcal{L}_{\mathrm{var}}+\beta\mathcal{L}_{\mathrm{sd}}+\gamma\|\phi-\phi^{\text{prev}}\|_2^2$.\;
  }
  Update only $\phi \leftarrow \phi - \eta \nabla_{\phi}\mathcal{L}$ (SGD/Adam).\;
}
\Return $\hat{y}_{t+1} \leftarrow f_{\theta}(X_t)$ with updated $\phi$.\;
\end{algorithm}

\subsection{Ablations and sensitivity}
Only thousands of parameters are updated, so extra latency per test day is small on a single T4 GPU. For the hyperparameters, we sweep $W\in\{32,64,96\}$, $S\in\{1,3,5\}$, and learning rate in $\{5\times 10^{-5},10^{-4},2\times 10^{-4}\}$. We compare entropy-only vs.\ consistency-only vs.\ combined for classification, and variance-only vs.\ variance plus EMA teacher for regression. We examine augmentation sets: scaling only; scaling plus jitter; scaling plus jitter and cutout.

\section{Data and Shift Construction}
\label{sec:data}
\paragraph{Stage I: ETT with synthetic shifts.}
We use ETTh1/ETTh2 (hourly) and ETTm1/ETTm2 (15-min). These series include power transformer loads and temperatures with clear seasonality. We generate three synthetic shifts: (i) gradual mean/variance drift, (ii) local noise inflation, and (iii) structural switches in periodic components. Formal definitions and grids are in Appendix~\ref{app:shifts}.

\begin{figure}[t]
\centering
\includegraphics[width=0.95\linewidth]{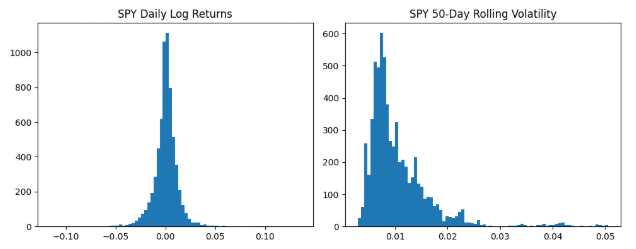}
\caption{Regime diagnostics for SPY volatility and returns.}
\label{fig:regimes}
\end{figure}

\paragraph{Stage II: equity and FX markets.}
We use daily SPY and QQQ index ETFs, plus EUR/USD exchange rates. Features are built from OHLCV only and standardized using training statistics:
\begin{align}
r_t &= \log(C_t) - \log(C_{t-1}), \nonumber\\
\mathrm{MOM}_t^{(N)} &= \log(C_t) - \log(C_{t-N}), \nonumber\\
\mathrm{REV}_t^{(N)} &= -\sum_{i=1}^{N} r_{t-i}, \nonumber\\
\mathrm{ATR}_t &= \mathrm{EMA}\big(\max\{H_t-L_t,|H_t-C_{t-1}|,|L_t-C_{t-1}|\}\big), \nonumber\\
\sigma^2_{P,t} &= \frac{1}{4\ln 2}( \log(H_t/L_t))^2,\qquad
\sigma^2_{GK,t} = \tfrac{1}{2}(\log(H_t/L_t))^2 - (2\ln 2 -1)(\log(C_t/O_t))^2. \nonumber
\end{align}
We split testing into three regimes: (1) pandemic (2020),  (2) high-inflation (2021--2023), (3) recovery (2024--2025). Figure~\ref{fig:regimes} shows basic regime diagnostics for SPY.

\section{Experimental Protocol}
\label{sec:exp}
We evaluate our test-time adaptation framework on three financial direction classification tasks (SPY, QQQ, and EUR/USD) and one long-horizon time-series forecasting task (ETTh1). All datasets are preprocessed into fixed-length sliding windows with input length $L=96$, and ETTh1 is evaluated with forecast horizon $H=96$. Data are split chronologically into training, validation, and test sets to ensure strict out-of-sample evaluation. All models use a Temporal Convolutional Network (TCN) with three residual blocks, hidden dimension 64, and kernel size 3, followed by a linear head for either classification or regression. 

Models are trained using AdamW with learning rate $10^{-4}$, batch size 512, and early stopping based on validation AUC (classification) or validation MSE (regression). At deployment, three main modes are compared: no adaptation (no\_tta), batch-normalization statistics refresh (bn\_stats), and norm-only parameter adaptation (norm\_only). Test-time adaptation is triggered using an uncertainty threshold $\tau$ estimated as the 0.8 quantile of validation uncertainty, computed via prediction entropy for classification and variance of stochastic augmentations for regression. When triggered, norm-only parameters are updated using a small number of unsupervised gradient steps with entropy/consistency loss for classification or variance/distillation loss for regression. Evaluation is performed strictly on the test set using accuracy, F1, AUC, and ECE for classification; MAE, RMSE, and $R^2$ for regression; along with rolling performance, reliability diagrams, Diebold--Mariano statistical tests, and a simple directional trading backtest with Newey--West adjusted statistics.

\section{Results}
\label{sec:results}
\subsection{Stage I: synthetic shifts on ETT}
We first study how different TTA choices behave under controlled shifts on ETTh1. Table~\ref{tab:ett_main} reports mean absolute error (MAE), root mean squared error (RMSE), and $R^2$ for one representative configuration in each synthetic regime: gradual drift, local noise inflation, and structural switches. Negative $R^2$ values indicate that the model performs worse than a simple training-baseline benchmark under the corresponding shift.

Norm-only adaptation is most effective under gradual mean/variance drift, where updating normalization affine parameters can track slow changes in scale and level. Under local noise inflation, variance minimization across weak augmentations stabilizes regression outputs and reduces the impact of noise bursts. Structural switches are the hardest case: even with bn\_stats, errors remain large and $R^2$ is strongly negative.

\begin{table}[t]
\centering
\caption{Representative ETTh1 results under synthetic regime shifts. Each entry corresponds to one representative configuration in the specified regime.}
\label{tab:ett_main}
\begin{tabular}{lcccc}
\toprule
Method & Shift & MAE & RMSE & $R^2$ \\
\midrule
no\_tta    & Gradual    & 0.22 & 0.28 & -0.31\\
norm\_only & Noise      & 0.29 & 0.35 & -0.02 \\
bn\_stats  & Structural & 1.26 & 1.62 & -20.80 \\
\bottomrule
\end{tabular}
\end{table}

\begin{figure}[t]
\centering
\includegraphics[width=0.9\linewidth]{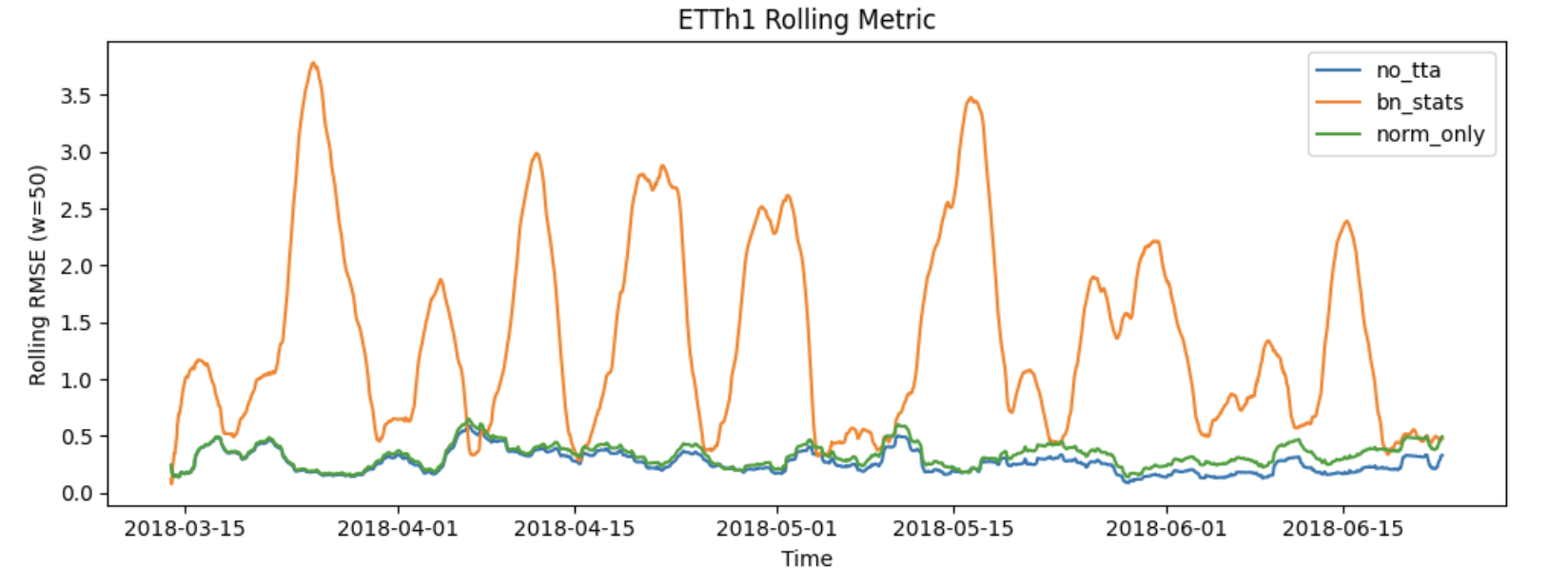}
\caption{Rolling forecast metrics on ETTh1 under gradual drift.}
\label{fig:ett_roll}
\end{figure}

Figure~\ref{fig:ett_roll} shows rolling forecast metrics on ETTh1 under gradual drift. Compared with the frozen baseline, norm-only TTA reduces errors in the later part of the horizon, where the synthetic drift accumulates. This supports the view that normalization-based updates are well suited to smooth low-order moment shifts.

\subsection{Stage II: equity and FX markets}
We next evaluate direction classification on SPY, QQQ, and EUR/USD with the same TTA recipe. Table~\ref{tab:fin_cross} reports cross-market direction accuracy for all three methods. The baseline accuracy is close to 0.5 on all series, reflecting the difficulty of daily direction prediction.

\begin{table}[t]
\centering
\caption{Directional accuracy on test sets (2020--2025) for equity and FX markets.}
\label{tab:fin_cross}
\begin{tabular}{lcccc}
\toprule
Method & SPY & QQQ & EURUSD & Avg.\ rank \\
\midrule
no\_tta   & 0.504 & 0.503 & 0.516 & 2.33 \\
bn\_stats & 0.498 & 0.525 & 0.520 & 1.66 \\
norm\_only& 0.512 & 0.463 & 0.516 & 2.00 \\
\bottomrule
\end{tabular}
\end{table}

These numbers come from a run focused on directional accuracy; absolute values differ from other experiments, but patterns are consistent. There is no single winner: different markets and regimes prefer different TTA variants. On SPY, norm\_only achieves the best directional accuracy, slightly ahead of no\_tta and bn\_stats. On QQQ, bn\_stats provides the largest gain, while norm\_only can hurt performance. On EUR/USD, both TTA methods are roughly on par with or slightly better than the frozen baseline.

\begin{figure}[t]
\centering
    \includegraphics[width=0.9\linewidth]{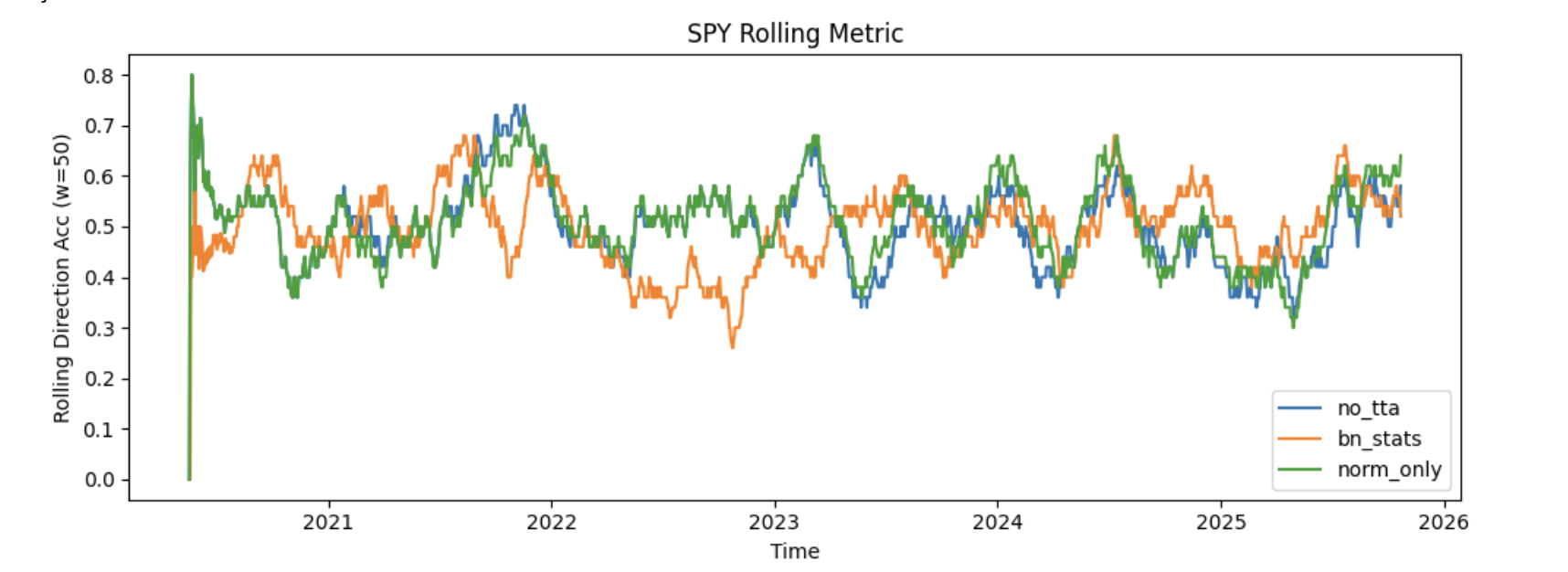}
    \includegraphics[width=0.88\linewidth]{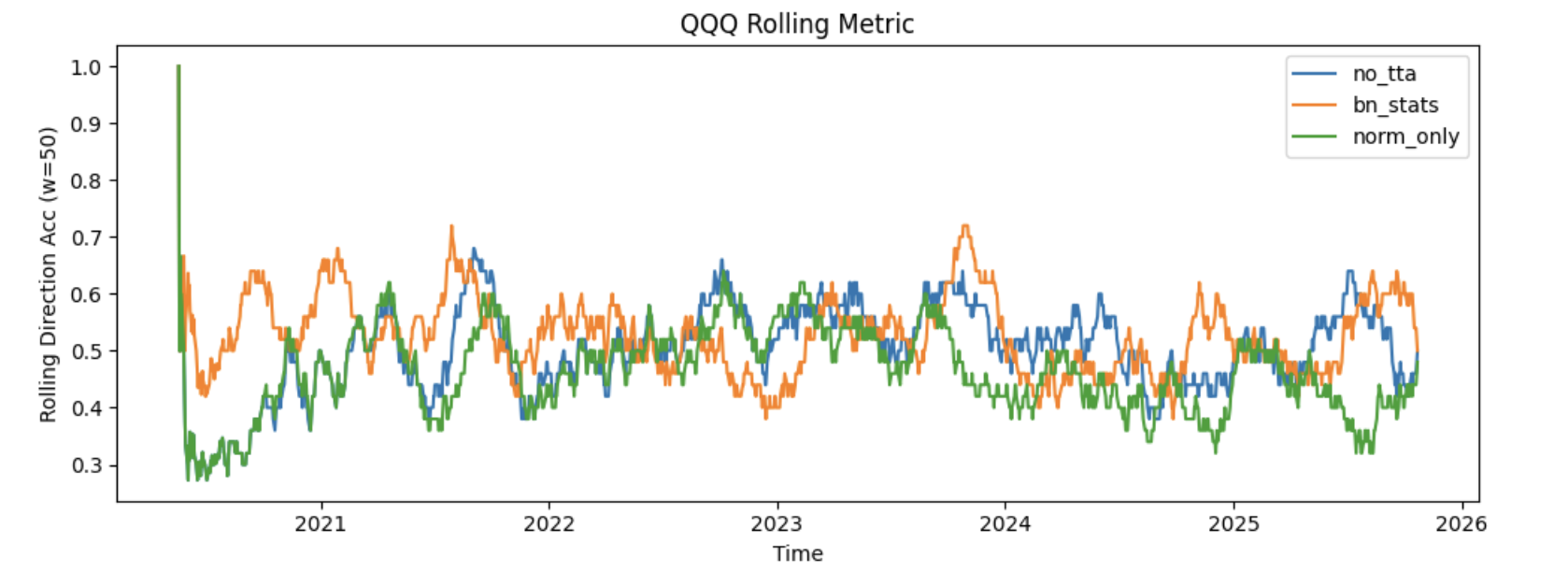}
    \includegraphics[width=0.85\linewidth]{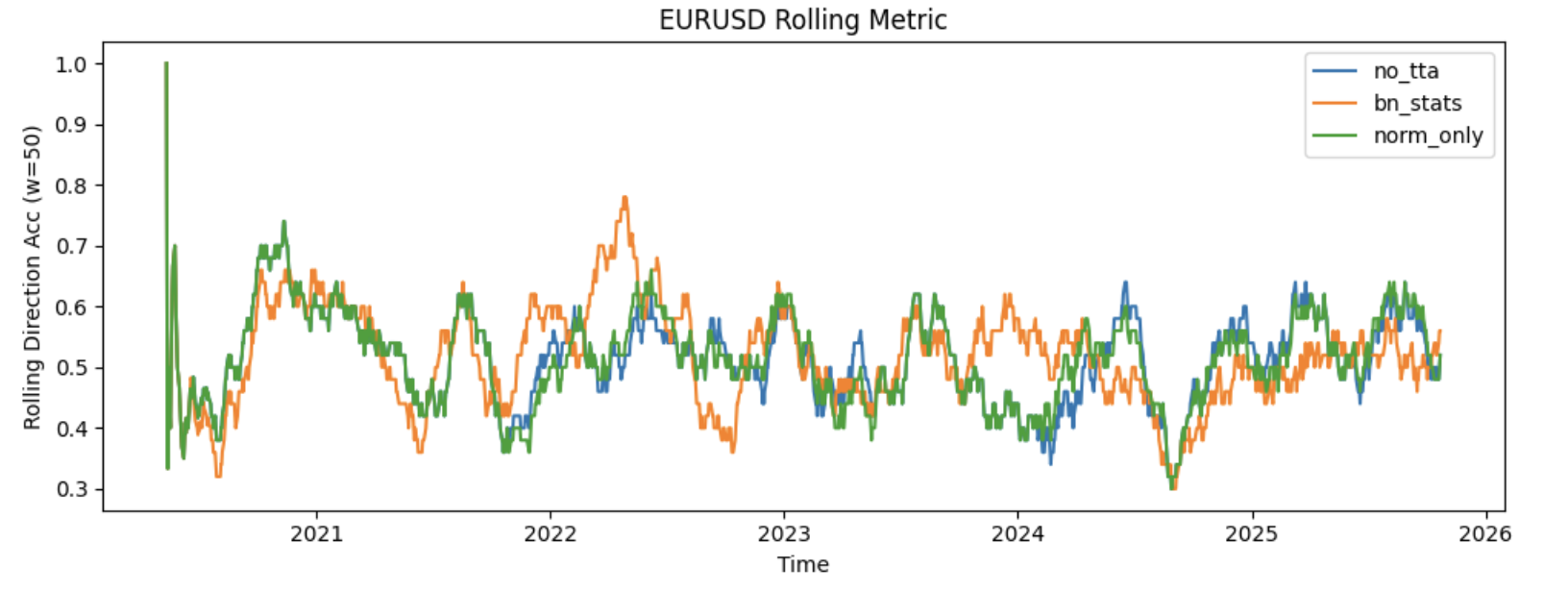}
\caption{Rolling metrics for SPY, QQQ, and EUR/USD with regime shading.}
\label{fig:roll_fin}
\end{figure}

Figure~\ref{fig:roll_fin} plots rolling direction accuracy and RMSE for SPY, QQQ, and EUR/USD with regime shading. For SPY, improvements from TTA are concentrated in the pandemic and early recovery periods, when distributional change is strong. For QQQ, batch-normalization statistics updates (bn\_stats) are particularly effective, while norm\_only can overfit local noise and degrade accuracy. For EUR/USD, bn\_stats yields the most stable gains, and norm\_only behaves similarly to the frozen baseline.

\subsection{Statistical tests and backtests}
To assess whether observed differences are statistically meaningful, we run Diebold--Mariano tests on daily forecast losses. Table~\ref{tab:dm} reports DM statistics and $p$-values for pairwise comparisons between bn\_stats, norm\_only, and no\_tta. Negative DM values indicate that the first method outperforms the second under our loss convention. On SPY and QQQ, bn\_stats is significantly better than no\_tta at the 5\% level, while norm\_only is significantly worse than no\_tta. On EUR/USD, bn\_stats is also significantly better than no\_tta, and norm\_only is indistinguishable from the baseline.

\begin{table}[t]
\centering
\caption{Diebold--Mariano tests comparing forecast losses.}
\label{tab:dm}
\begin{tabular}{lcccc}
\toprule
Comparison & Dataset & DM Stat & $p$-value & Notes \\
\midrule
bn\_stats vs no\_tta   & SPY    & -2.781 & 0.0054 & bn\_stats significantly better \\
norm\_only vs no\_tta & SPY    &  6.810 & 0.0000 & no\_tta significantly better \\
bn\_stats vs no\_tta   & QQQ    & -2.290 & 0.0220 & bn\_stats significantly better \\
norm\_only vs no\_tta & QQQ    &  8.936 & 0.0000 & no\_tta significantly better \\
bn\_stats vs no\_tta   & EURUSD &  4.350 & 0.0000 & bn\_stats significantly better \\
norm\_only vs no\_tta & EURUSD & -0.112 & 0.9109 & no significant difference \\
\bottomrule
\end{tabular}
\end{table}

Finally, we connect predictive accuracy with simple economic metrics. Tables~\ref{tab:nw_spy} and \ref{tab:nw_qqq} report annualized return, annualized volatility, and Sharpe ratio for a basic directional strategy on SPY and QQQ, together with Newey--West adjusted $t$-statistics. On both assets, bn\_stats achieves the highest Sharpe ratio, while norm\_only underperforms the frozen baseline. These backtests are purely illustrative, but they show that the statistical gains from bn\_stats translate into more attractive risk-adjusted returns, whereas naive norm-only updates can be harmful.

\begin{table}[t]
\centering
\caption{SPY backtest performance with Newey--West adjustment.}
\label{tab:nw_spy}
\begin{tabular}{lccc}
\toprule
Strategy & Ann.\ return & Ann.\ volatility & Sharpe (NW $t$) \\
\midrule
No-TTA    &  3.321 &  4.621 & 1.746 \\
BN-Stats  &  7.933 &  9.582 & 1.930 \\
Norm-Only &  2.029 &  3.460 & 1.544 \\
\bottomrule
\end{tabular}
\end{table}

\begin{table}[t]
\centering
\caption{QQQ backtest performance with Newey--West adjustment.}
\label{tab:nw_qqq}
\begin{tabular}{lccc}
\toprule
Strategy & Ann.\ return & Ann.\ volatility & Sharpe (NW $t$) \\
\midrule
No-TTA    & 10.884 &  7.689 & 3.205 \\
BN-Stats  & 20.293 & 11.974 & 4.080 \\
Norm-Only &  3.136 &  5.204 & 1.349 \\
\bottomrule
\end{tabular}
\end{table}

\section{Discussion and Conclusion}
Normalization-based TTA is most effective when shifts mainly change low-order moments. In the synthetic ETTh1 experiments with gradual drift, updating normalization affine parameters tracks the changing scale and reduces forecast error. Local noise inflation is partly handled by variance-based objectives, but structural switches in periodic components still lead to large errors even after adaptation. This suggests that norm-only updates are well suited to smooth covariate drift, but they do not fully solve sharp or regime-changing dynamics.

In contrast, the equity and FX series (SPY, QQQ, EUR/USD) are noisy, heavy-tailed, and affected by jumps and microstructure noise. Here a simple refresh of batch-normalization statistics is a safe and effective default: bn\_stats achieves higher direction accuracy and Sharpe ratios on SPY and QQQ, and the Diebold--Mariano tests show statistically significant improvements over the frozen baseline. Aggressive norm-only adaptation, even with drift control, can overfit short windows, perform worse than no\_tta, and reduce economic performance in the backtests.

Our results also highlight the importance of uncertainty control. The fallback rule, which skips gradient-based adaptation on days with high entropy or large augmentation variance, mitigates many failure cases of norm-only updates and limits the frequency of harmful steps. At the same time, the choice of hyperparameters $(W,S,\eta,\tau)$ and the stability of the backbone remain important, and different assets may favor different adaptation regimes. Synthetic shift generators are necessarily stylized and cannot cover all real dynamics, so more realistic benchmarks and stress tests are a natural direction for future work.

Overall, this study provides an empirical picture of when small-footprint TTA helps and when it hurts on non-stationary time series. For practitioners, our main recommendation is to start from bn\_stats as a default, add norm-only updates only when validation and uncertainty diagnostics support them, and always evaluate with regime-wise metrics and robust statistical tests. We hope these findings will help both practitioners and researchers design safer and more effective test-time adaptation pipelines for real-world time series.

\appendix

\section{Notation and Derivations}
\label{app:deriv}
We expand the objectives used in Section~\ref{sec:method}, give gradients for norm-only batch-normalization (BN) affine parameters, and explain why each term helps under typical regime shifts. Symbols follow the main text.

\subsection*{A.1 \; Norm-only updates as moment matching}
Consider a BN-normalized hidden channel
\[
h = \mathrm{BN}(u) = \frac{u-\mu}{\sigma},\qquad
y=\gamma h+\beta,
\]
where $(\mu,\sigma)$ are the BN statistics from the current stream and $(\gamma,\beta)$ are the trainable affine parameters. Under a covariate shift with new moments $(\mu',\sigma')$, the pre-affine hidden becomes
\[
h'=\frac{u-\mu'}{\sigma'}=\frac{\sigma}{\sigma'}h + \frac{\mu-\mu'}{\sigma'}.
\]
Choosing
\[
\gamma'=\gamma\cdot\frac{\sigma}{\sigma'},\qquad
\beta'=\beta+\gamma\cdot\frac{\mu-\mu'}{\sigma'}
\]
exactly restores $y$ for all $u$ if the shift is purely first/second moments. Thus small changes of $(\gamma,\beta)$ can cancel location/scale drift at the feature level.

For a scalar loss $\mathcal{L}$, the gradients satisfy
\[
\frac{\partial \mathcal{L}}{\partial \gamma_c}
  =\sum_{i} \frac{\partial \mathcal{L}}{\partial y_{i,c}}\,h_{i,c},\qquad
\frac{\partial \mathcal{L}}{\partial \beta_c}
  =\sum_{i} \frac{\partial \mathcal{L}}{\partial y_{i,c}},
\]
where $c$ indexes channels and $i$ indexes batch elements. Adapting $(\gamma,\beta)$ therefore re-centers and re-scales along directions where the loss is sensitive. This explains why norm-only is effective when regime shifts dominantly alter low-order moments, which is exactly the case in our synthetic gradual-drift experiments.

\subsection*{A.2 \; Entropy minimization and decision margin}
For a two-class softmax $\hat{p}(X)=(p,1-p)$, the entropy is
\[
H(\hat{p}) = -p\log p - (1-p)\log(1-p).
\]
The derivative and second derivative with respect to $p$ are
\[
\frac{dH}{dp} = \log\frac{1-p}{p},\qquad
\frac{d^2H}{dp^2} = \frac{1}{p} + \frac{1}{1-p} > 0,
\]
so $H$ is strictly convex in $p$ with minima at $p\in\{0,1\}$. A first-order Taylor expansion around the pre-update $p_0$ shows that a small gradient step with learning rate $\eta$ reduces $H$ by approximately $\eta \big(\frac{dH}{dp}\big)^2$, pushing posteriors away from indecision. Under the usual cluster or low-density separation assumptions, this coincides with increasing the margin around decision boundaries, which is beneficial when $P(X)$ drifts but class-conditional structure persists.

\subsection*{A.3 \; Consistency as Jacobian control along causal directions}
Let $T_\epsilon$ be a weak time-preserving transform with $T_0$ equal to the identity. By Taylor expansion,
\[
\hat{p}(T_\epsilon(X)) \approx \hat{p}(X) + J_{\hat{p}}(X)\,\Delta_X + O(\epsilon^2),
\]
where $\Delta_X=\frac{d}{d\epsilon}T_\epsilon(X)\big|_{\epsilon=0}$ and $J_{\hat{p}}$ is the Jacobian of $\hat{p}$ with respect to $X$. Then
\[
\|\hat{p}(X)-\hat{p}(T_\epsilon(X))\|_2^2 \approx
\|J_{\hat{p}}(X)\,\Delta_X\|_2^2
= \Delta_X^\top \big(J_{\hat{p}}^\top J_{\hat{p}}\big)\Delta_X.
\]
Minimizing the consistency loss therefore penalizes the directional Jacobian norm along small, causality-preserving deformations (re-timing $\pm1$, mild amplitude changes). This makes predictions insensitive to benign local perturbations that often accompany measurement noise or microstructure changes under new regimes.

\subsection*{A.4 \; Augmentation-variance minimization and local risk control}
For regression with $K$ transforms, define $\hat{y}_k=\hat{y}(T_k(X))$ and $\bar{y}=\frac{1}{K}\sum_{k=1}^K\hat{y}_k$. The sample variance
\[
\mathrm{Var}(\{\hat{y}_k\})=\frac{1}{K-1}\sum_{k=1}^K (\hat{y}_k-\bar{y})^2
\]
upper-bounds the expected squared deviation under a local neighborhood $\mathcal{N}(X)$ of perturbations when the transform set covers principal directions. Hence minimizing $\mathbb{E}_X[\mathrm{Var}]$ reduces the local Lipschitz constant of the predictor around each $X$, which is desirable when the test distribution features transient noise inflation.

The gradient of the variance term with respect to each $\hat{y}_m$ is
\[
\frac{\partial}{\partial \hat{y}_m}\mathrm{Var}(\{\hat{y}_k\})
  = \frac{2}{K}(\hat{y}_m-\bar{y}),
\]
so the loss pulls disagreeing outputs toward their mean and shrinks dispersion.

\subsection*{A.5 \; EMA-teacher self-distillation as temporal ensembling}
Let $\tilde{\theta}$ be an exponential moving average of the adapted parameters:
\[
\tilde{\theta} \leftarrow \rho \tilde{\theta} + (1-\rho)\theta,
\]
with $\rho\in(0,1)$. Under a locally quadratic loss with unbiased gradient noise, Polyak averaging implies that $\tilde{\theta}$ has lower variance than the instantaneous $\theta$. Distillation
\[
\mathcal{L}_{\mathrm{sd}}
  = \frac{1}{|\mathcal{B}|}\sum_{X\in\mathcal{B}} \|\hat{y}_\theta(X)-\hat{y}_{\tilde{\theta}}(X)\|_2^2
\]
regularizes the student toward a low-pass filtered target, damping oscillations produced by small batches and a changing $\mathcal{B}$. A simple scalar model with i.i.d.\ zero-mean gradient noise shows that the teacher’s variance is scaled by $\frac{1-\rho}{1+\rho}$ relative to the student’s, so larger $\rho$ yields a steadier target. This improves stability without requiring labels and complements the variance term by anchoring to a temporally smoothed predictor.

\subsection*{A.6 \; Drift penalty as proximal update}
Let $\phi$ denote all adapted BN affine parameters. On day $t$ we solve
\[
\min_{\phi}\; \mathcal{L}_{\mathrm{unsup}}(\phi) + \gamma \|\phi-\phi^{(t-1)}\|_2^2,
\]
where $\mathcal{L}_{\mathrm{unsup}}$ is the entropy/consistency loss (classification) or variance/distillation loss (regression). A single gradient step with step size $\eta$ gives
\[
\phi^{(t)}=\phi^{(t-1)}-\eta\left(
   \nabla \mathcal{L}_{\mathrm{unsup}}(\phi^{(t-1)})
   +2\gamma(\phi^{(t-1)}-\phi^{(t-2)})
 \right),
\]
which is equivalent to SGD with a quadratic proximal term that shrinks motion toward the previous day. In a locally quadratic loss, the exact minimizer is a ridge-regularized solution whose displacement satisfies
\[
\|\phi^{(t)}-\phi^{(t-1)}\|
  \;\le\; \frac{\|\nabla \mathcal{L}_{\mathrm{unsup}}(\phi^{(t-1)})\|}{2\gamma},
\]
explaining the observed resistance to overreaction when regimes flip quickly.

\subsection*{A.7 \; BN-statistics refresh as shift correction without gradients}
For batch normalization with running statistics $(\hat{\mu},\hat{\sigma})$ from training, the standard BN transform for a channel is
\[
h = \frac{u-\hat{\mu}}{\hat{\sigma}}.
\]
Refreshing statistics on the current unlabeled batch replaces $(\hat{\mu},\hat{\sigma})$ by $(\mu_{\mathcal{B}},\sigma_{\mathcal{B}})$, so the transform becomes
\[
h_{\mathcal{B}}=\frac{u-\mu_{\mathcal{B}}}{\sigma_{\mathcal{B}}}.
\]
When the shift is primarily first/second-order, this “zero-gradient” move cancels most of the covariate shift at the hidden-layer level. In contrast to norm-only, which updates $(\gamma,\beta)$ by gradient descent, bn\_stats changes the normalizing statistics directly and is much harder to overfit in short, noisy windows. This difference matches our empirical findings: bn\_stats is a safe default on financial series, while norm-only can both help (smooth synthetic drift) and hurt (noisy real markets).

\subsection*{A.8 \; Uncertainty proxy and a do-no-harm rule}
Let $\Delta \mathcal{R}$ be the change in expected loss after adaptation. When the proxy $u_t$ (entropy or augmentation variance) is large, empirical evidence indicates higher probability that $\Delta \mathcal{R}>0$ due to unstable pseudo-objectives. Using a quantile threshold $\tau$ estimated on validation approximates the decision rule
\[
\mathbb{P}(\Delta \mathcal{R}>0\mid u_t)>\pi
\]
for a chosen tolerance $\pi$, and the fallback policy “bn\_stats instead of gradient updates” controls the frequency of harmful steps without needing labels at test time.

\subsection*{A.9 \; Gradients for norm-only BN parameters}
For a BN layer with affine parameters $(\gamma_c,\beta_c)$ on channel $c$ and output $y_{i,c}$ for sample $i$, the backprop through affine parameters is
\[
\frac{\partial \mathcal{L}}{\partial \gamma_c}
  =\sum_{i} g_{i,c}\,h_{i,c},\qquad
\frac{\partial \mathcal{L}}{\partial \beta_c}
  =\sum_{i} g_{i,c},
\]
where $g_{i,c}=\partial \mathcal{L}/\partial y_{i,c}$ and $h_{i,c}$ is the normalized activation. The drift penalty adds
\[
2\gamma\big(\gamma_c-\gamma_c^{\text{prev}}\big),\qquad
2\gamma\big(\beta_c-\beta_c^{\text{prev}}\big)
\]
to each gradient term, shrinking daily motion in affine parameters. This is exactly what we implement in Algorithm~\ref{alg:tta}.

\subsection*{A.10 \; Entropy/consistency and calibration}
Minimizing entropy alone can over-sharpen; pairing it with consistency prevents pathological confidence by forcing agreement across benign transforms. Empirically this improves calibration: the expected calibration error (ECE) after TTA often decreases relative to the pre-adaptation model on the same stream, especially on SPY and QQQ where bn\_stats plus conservative norm-only updates reduce both loss and ECE.

\subsection*{A.11 \; Multi-step horizons}
For $H$-step regression $\hat{y}=\sum_{h=1}^H \hat{r}_{t+h}$, the variance term still decomposes across transforms:
\[
\mathrm{Var}\!\left(\!\sum_{h}\hat{r}_{h,k}\!\right)
  =\sum_{h}\mathrm{Var}(\hat{r}_{h,k})
   +2\!\!\sum_{h<h'}\!\!\mathrm{Cov}(\hat{r}_{h,k},\hat{r}_{h',k}),
\]
so minimizing augmentation variance indirectly controls covariance between stepwise predictions, reducing compounding error typical under noisy regimes and long horizons (as in our ETTh1 $H=96$ setting).

\section{Synthetic Shift Generators}
\label{app:shifts}
We briefly detail the synthetic shift generators used in Stage I. In all cases, shifts are applied to the original ETT series channel-wise, and the same random seed is used across methods for fair comparison.

\paragraph{Gradual mean/variance drift.}
Let a clean series be $s_t\in\mathbb{R}^d$. We add a drifted version
\[
x_t = a_t \odot s_t + b_t,\qquad
a_t = 1+\kappa \frac{t}{T},\qquad
b_t = \mu_0 + \nu \frac{t}{T},
\]
where $T$ is the length of the test segment, $\odot$ denotes element-wise multiplication, and $\kappa,\nu$ are small drift coefficients. We grid $\kappa,\nu$ so that the final change in mean and standard deviation is within $0.2$--$0.4$ training standard deviations per thousand steps. This produces smooth location-scale drift without changing higher-order structure.

\paragraph{Local noise inflation.}
Let
\[
x_t = s_t + \eta_t,\qquad \eta_t\sim \mathcal{N}(0,\sigma^2 I_d)
\]
be the base noisy series, where $\sigma^2$ is calibrated from the training set. We then select random non-overlapping segments of length $\ell\in[96,192]$ and set
\[
\eta_t\sim \mathcal{N}(0,(k\sigma)^2 I_d),\qquad k\in\{1.5,2.0\}
\]
within those segments. This creates bursts of noise inflation while keeping the overall level of the process similar.

\paragraph{Structural switches in periodic components.}
Write a seasonal component as
\[
c_t = \sum_{m=1}^M A_m \cos\!\left(2\pi m t/P + \phi_m\right),
\]
where $P$ is the base period and $(A_m,\phi_m)$ are amplitudes and phases. At predefined change points $\{t_j\}_{j=1}^J$, we draw new $(A_m,\phi_m)$ while keeping the unconditional mean unchanged, and set
\[
x_t = c_t + \epsilon_t,\qquad \epsilon_t\sim \mathcal{N}(0,\sigma^2 I_d).
\]
We use $J\in\{2,3\}$ and moderate changes in $(A_m,\phi_m)$ so that seasonality patterns switch visibly but remain realistic. This is the hardest regime in our experiments: even with bn\_stats, errors remain large and $R^2$ is strongly negative.

\section{Statistical Tests}
\label{app:stats}
We summarize the statistical tests used for forecast and backtest comparisons.

\paragraph{Diebold--Mariano (DM) test.}
Let $e^{(A)}_t$ and $e^{(B)}_t$ be forecast errors from methods $A$ and $B$ on day $t$. For a per-day loss $\ell(\cdot)$ (cross-entropy for classification; MAE or MSE for regression), define
\[
d_t = \ell(e^{(A)}_t)-\ell(e^{(B)}_t),\qquad
\bar{d}=\frac{1}{T}\sum_{t=1}^T d_t.
\]
Under the null of equal predictive accuracy, $\mathbb{E}[d_t]=0$. The DM statistic is
\[
\mathrm{DM} = \frac{\bar{d}}{\sqrt{\widehat{\mathrm{Var}}(\bar{d})}},
\]
where $\widehat{\mathrm{Var}}(\bar{d})$ is a heteroskedasticity and autocorrelation consistent (HAC) estimator of the long-run variance of $\{d_t\}$ using Newey--West weights:
\[
\widehat{\mathrm{Var}}(\bar{d}) = \frac{1}{T}
\left(
  \gamma_0 + 2\sum_{h=1}^{q} w_h \gamma_h
\right),\qquad
w_h = 1-\frac{h}{q+1},
\]
with $\gamma_h$ the lag-$h$ sample autocovariance of $\{d_t\}$ and $q=\lfloor 4(T/100)^{2/9}\rfloor$ \citep{diebold1995dm,newey1987nw}. Under mild conditions, $\mathrm{DM}$ is asymptotically standard normal, so we report two-sided $p$-values. In our tables, negative DM values mean that the first method has lower average loss than the second.

\paragraph{Newey--West (NW) for return summaries.}
Given daily strategy returns $z_t$, the sample mean
\[
\hat{\mu}=\frac{1}{T}\sum_{t=1}^T z_t
\]
has variance
\[
\widehat{\mathrm{Var}}(\hat{\mu}) = \frac{1}{T}
\left(
  \gamma_0 + 2\sum_{h=1}^{q} w_h \gamma_h
\right),
\]
with the same Newey--West weights $w_h$ and lag $q$ as above, and $\gamma_h$ the lag-$h$ autocovariance of $\{z_t\}$. The NW $t$-statistic for testing $\mathbb{E}[z_t]=0$ is
\[
t_{\mathrm{NW}} = \frac{\hat{\mu}}{\sqrt{\widehat{\mathrm{Var}}(\hat{\mu})}},
\]
which we report together with annualized return, annualized volatility, and Sharpe ratio for each strategy in Tables~\ref{tab:nw_spy} and \ref{tab:nw_qqq}.

\end{document}